\documentclass[nofootinbib,twocolumn,aps,amsmath,amssymb]{revtex4}
\usepackage{graphics,bm}
\usepackage{epsfig}
\usepackage{graphicx}
\usepackage{dcolumn}
\newcommand{\beq}{\begin{equation}}
\newcommand{\eeq}{\end{equation}}
\newcommand{\bqa}{\begin{eqnarray}}
\newcommand{\eqa}{\end{eqnarray}}
\newcommand{\la}{\langle}
\newcommand{\ra}{\rangle}
\newcommand{\amp}[1]{\la #1 \ra}
\newcommand{\spaceint}[2]
{
 \frac{\mathrm{d}^{#1}{#2}}{(2 \pi)^{#1}}
}	
\newcommand{\li}{\,\mathrm{li}\,}                       
\newcommand{\measure}[1]{\mathrm{d}#1\,}		
\newcommand{\simordertwo}{\raisebox{-4pt}{$\, \stackrel{\textstyle <}{\sim}
\,$}}
\begin{document}

\parindent=20pt
\parskip=0pt
\def\sumint{\hbox{$\sum$}\!\!\!\!\!\!\int}

\title{Thermodynamics of the $O(N)$ Nonlinear Sigma Model in 1+1 Dimensions}
\author{Jens O. Andersen}
\affiliation{Nordita,\\
Blegdamsvej 17, 2100 Copenhagen, Denmark}
\author{Dani\"el Boer and Harmen J. Warringa}
\affiliation{Department of Physics and Astronomy, Vrije Universiteit, 
De Boelelaan 1081, 1081 HV Amsterdam, The Netherlands}
\date{\today}

\begin{abstract}
The thermodynamics of the $O(N)$ nonlinear sigma model in 1+1 dimensions 
is studied. We calculate the pressure to next-to-leading order in the 
$1/N$ expansion and show that at this order, only the minimum 
of the effective potential can be rendered finite by temperature-independent 
renormalization. To obtain a finite effective potential away from the minimum 
requires an arbitrary choice of prescription, which implies that the 
temperature dependence is ambiguous. We show that the
problem is linked to thermal infrared renormalons.
\end{abstract}

\maketitle

\section{Introduction}

The $O(N)$ nonlinear sigma model (NLSM) in 1+1 dimensions has been studied
extensively at zero temperature as a toy model for QCD.
It is a remarkably rich theory, which is asymptotically free and has a
dynamically generated mass gap. It is renormalizable both perturbatively and 
in the $1/N$ expansion. Moreover, for $N=3$ it has instanton solutions. 
Unlike the NLSM in more than two dimensions, where the theory is no longer 
renormalizable, there is no spontaneous symmetry breaking of the global 
$O(N)$ symmetry. This reflects the Mermin-Wagner-Coleman theorem
\cite{Mermin-Wagner,coleman}, 
which
forbids spontaneous breakdown of a continuous symmetry
in a homogeneous system in one spatial dimension at 
any temperature.
Moreover, the model suffers from infrared (IR) divergences 
in perturbation theory, since the fields are massless in that 
case~\cite{Zinn-Justin}. 
However, it was conjectured by Elitzur~\cite{elit1} and shown by  
David~\cite{David-81} that the infrared divergences cancel in 
$O(N)$-invariant correlation functions. In addition, a mass gap is generated 
nonperturbatively. In the large-$N$ limit, which is equivalent to summing all 
so-called daisy and superdaisy graphs, $m=\mu\exp{(-2\pi/g^2)}$, where
$g$ is the coupling constant and $\mu$ is the renormalization scale.

Dine and Fischler~\cite{dine} investigated the NLSM in $1+1$ dimensions at 
finite temperature. They calculated the free energy in perturbation theory 
and in the large-$N$ limit.
In the weak-coupling expansion, they showed that the two-loop contribution to
the ideal gas vanishes and that the three-loop contribution is
infrared finite; the latter in fact also vanishes~\cite{future}.
The leading-order calculation in the $1/N$ 
expansion shows that a thermal mass of order $Ng^2T$ arises. This is a 
nonperturbative result that shows that one is effectively dealing with a gas 
of massive particles. 

In this paper, we extend the analysis of Dine and Fischler to 
next-to-leading-order (NLO) in the $1/N$ expansion. At zero temperature, the
effective potential (or equivalently the Gibbs free energy) has been 
investigated at this order by Biscari {\it et al.}~\cite{BCR}. 
The $1/N$ correction to 
the thermodynamic potential has the interesting feature of containing
renormalon singularities. We will show that it cannot in general 
be renormalized
in a temperature-independent way, except at its minimum as 
a function of $m^2$. Away from the
minimum, one will need to introduce a temperature-dependent prescription to
deal with the poles in the Borel plane.
 
Much is known about IR renormalons in the $O(N)$ NLSM 
in 1+1 dimensions~\cite{David-82,David-84,David-86,Beneke,Beneke-Braun},
but the consequences at finite
temperature have not yet been investigated. 
Thermal renormalons have been studied by Loewe and Valenzuela \cite{Loewe}
in $\phi^4$ theory in 3+1 dimensions. In this theory, one deals with 
ultraviolet (UV)
renormalons only and thus it resembles QED rather than QCD.
They show that the residues of the UV
renormalon poles in the Borel 
plane (which are on the positive real axis, whereas in QCD they would be on
the negative real axis due to asymptotic freedom, such that they do not affect
the Borel transform~\cite{'tHooft:am}) in general  
are temperature dependent, but the positions of the poles are not. 
We will show that this is also the case for IR renormalons, except at the
minimum of the effective potential, where also the residues are temperature
independent. 

Blaizot {\it et al.}~\cite{Blaizot} have recently studied the 
Gross-Neveu model in 1+1 dimensions at finite temperature  
at NLO in the $1/N$ expansion.
While there are similarities between this model and the NLSM,
such as dynamical mass generation and asymptotic freedom, 
no problems related to IR
renormalons are encountered in Ref.~\cite{Blaizot} (see also~\cite{Kneur}).
One can uniquely define the effective potential at NLO at nonzero 
temperature. 

The paper is organized as follows. In Sec.~II, we discuss the NLSM
at zero temperature. In Sec.~III, we calculate the
finite-temperature pressure at NLO.
In Sec.~IV, we discuss various approximations and compare them with exact
numerical
results. In Sec.~V, we show that one cannot define an off-shell 
effective potential
and this is related to thermal infrared renormalons. In Sec.~VI, we summarize
and conclude. 

\section{Zero temperature}

The Euclidean Lagrangian for the nonlinear sigma model is
\bqa
{\cal L}&=&{1\over2}(\partial_{\mu}\Phi)^2 
+{1\over2}\alpha(\Phi^2-N g^{-2})\;,
\label{lagrang}
\eqa
where the scalar field $\Phi=(\phi_1,\phi_2...,\phi_N)$ 
forms an $N$-component vector and $\alpha$ is a Lagrange multiplier that
enforces the constraint $\Phi^2(x)=N g^{-2}$. 
The auxiliary field $\alpha$ is now written as 
the sum of a space-time independent background $m^2$
and a quantum fluctuating
field $\tilde{\alpha}$; $\alpha=m^2+\tilde{\alpha}$.
The Green's functions
of $\Phi$ require wavefunction and coupling constant renormalization
in the $1/N$ expansion,
as discussed by Rim and Weisberger~\cite{Rim-Weisberger} through NLO.
 
The Lagrangian in Eq.~(\ref{lagrang}) is quadratic in the fields $\Phi$
and the integral over $\Phi$ can therefore be done exactly. 
One then obtains (cf.\ e.g.\ \cite{Novikov})
\bqa\nonumber
{\cal Z}&=&\int{\cal D}\tilde \alpha\;\exp \Bigg\{
-{N\over2}\mbox{tr}\ln
\left[p^2+m^2+\tilde{\alpha}\right]+\\
&& N \int_0^{\beta }\mathrm{d}\tau \int \mathrm{d}x
\left[{1\over2}m^2g^{-2}+
{1\over2}\tilde{\alpha}g^{-2}
\right]
\Bigg\}\;,
\label{fd}
\eqa
where $\beta=1/T$, such that at zero temperature $\beta = \infty$.
The next step is to expand the functional determinant around the classical
solution $\tilde{\alpha}=0$ and integrate over $\tilde{\alpha}$. 
By scaling $\tilde{\alpha}\rightarrow\tilde{\alpha}/\sqrt{N}$, it is seen
that this expansion is equivalent to a $1/N$ expansion.

It is important to realize that $m^2$ is by definition the
vacuum expectation value of $\alpha$, i.e.\ it is the quantity with respect
to which we will minimize the effective potential
in order to obtain the pressure. 
Beyond leading order in the $1/N$ expansion $m^2$ receives divergent 
contributions. In order to show this, we rewrite the expression for $m^2$
in terms of $m_\phi^2$,
where $m_{\phi}$ is defined as the pole of the propagator
$D_\phi(P, m)$ of $\Phi$. At NLO, $D_\phi(P, m)$ is given by 
\begin{equation}
  D_\phi(P, m) = \frac{Z_\phi}{P^2 + m^2 - \frac{1}{N} \Sigma(P, m)}\;,
\end{equation}
where $Z_\phi$ is the wavefunction renormalization constant and
\begin{equation}
  \Sigma(P, m) = \int \spaceint{2}{Q} \frac{1}{(P+Q)^2 + m^2} \frac{1}{\Pi(Q,
  m)} \;
\end{equation}
is the self-energy function.
Here, $\Pi(P,m)$ is the inverse propagator for $\tilde \alpha$:
\bqa\nonumber 
\Pi(P,m) & = & -{1\over 2} \int\frac{\mathrm{d}^2 Q}{(2\pi)^2}
{1\over Q^2+m^2}{1\over(P+Q)^2+m^2}\;\\
& = & -\frac{1}{4\pi P^2 \xi} \ln \left(\frac{\xi+1}{\xi -1} \right)\;,
\label{Pi}
\eqa
where $\xi = \sqrt{1+4m^2/P^2}$. 
Choosing renormalization point $P^2=-m_\phi^2$, one obtains~\cite{flyvbjerg}, 
\beq
  m_\phi^2 = m^2 + \frac{m^2}{N} \li\left(\frac{\Lambda^2}{m^2}\right)\;,
\label{massshift}
\eeq
where $\mathrm{li}(x)$ is the logarithmic integral,
\begin{equation}
  \mathrm{li}(x) = \mathcal{P} \int^x_0 \measure{t} \frac{1}{\ln t} \;.
\end{equation}
Here, $\Lambda$ denotes the ultraviolet momentum cutoff and ${\cal P}$
indicates a principal-value prescription for the integral.
Solving Eq.~(\ref{massshift}) for $m^2$, we obtain
\beq
m^2 = m_\phi^2 - \frac{m_\phi^2}{N}\li\left(\frac{\Lambda^2}{m_\phi^2}
\right)\;.
\label{massshift2}
\eeq
In order to have a pole with a residue equal to unity, one needs 
$Z_\phi = 1 - \frac{1}{N} \Sigma'(P, m)|_{P^2=-m_\phi^2}$, which yields 
\beq
Z_\phi = 1 + \frac{1}{N} \ln \ln \left(\frac{\Lambda^2}{\mu^2}\right)\;.
\label{wf1}
\eeq
The wavefunction renormalization Eq.~(\ref{wf1}) 
is in accordance with that obtained by
\mbox{Flyvbjerg~\cite{flyvbjerg}}.
Rim and Weisberger \cite{Rim-Weisberger}
calculated the wavefunction renormalization constant 
in dimensional regularization and it also agrees with Eq.~(\ref{wf1})
as can be seen by identifying $\ln (\Lambda^2/\mu^2) \to2/\epsilon$ 
where $d=2-\epsilon$. 

The effective potential ${\cal V}$ through next-to-leading order in 
the $1/N$ expansion is given by 
\bqa\nonumber
{\cal V}&=& {m^2 N\over2g_b^{2}} - {1\over2}N
\int\frac{\mathrm{d}^2 P}{(2\pi)^2} 
\ln\left[P^2+m^2\right]
\\ &&
-{1\over2}\int \frac{\mathrm{d}^2 P}{(2\pi)^2}\ln\left[\Pi(P,m)\right]\;,
\label{V0}
\eqa
where we have added a subscript $b$ to $g$ to 
indicate explicitly that it is the bare coupling constant.
Evaluating the integrals in Eq.~(\ref{V0}) using an ultraviolet momentum 
cutoff $\Lambda$,
one obtains
\bqa\nonumber
{\cal V} & = & \frac{m^2 N}{2 g_b^2} - 
\frac{m^2 N}{8\pi}\left(1 + \ln \frac{\Lambda^2}{m^2} \right)
\\
\nonumber
& & - \frac{1}{8\pi} 
\left[\left(\Lambda^2 + 2 m^2\right) \ln \ln {\Lambda^2 \over m^2} \right. \\
&& \mbox{} \hspace{-1.2cm} 
\left. - m^2 {\rm li}\left({\Lambda^2\over m^2}\right) 
+ 2 m^2 \left(\gamma_E -1 -
\ln{\Lambda^2\over 4m^2} \right) \right]\;.
\label{intlogPi}
\eqa
Here and in the subsequent results we have dropped $m$-independent 
divergences and terms that vanish in the limit $\Lambda^2 \to \infty$.

To obtain the pressure, one evaluates the effective potential at its minimum.
The condition for the minimum is given by equation
\bqa
{\partial{\cal V}\over\partial m^2}&=&0\;.
\label{gap}
\eqa
Eq.~(\ref{gap}) is often referred to as a gap equation. 
Differentiating Eq.~(\ref{intlogPi}), one obtains
\bqa \nonumber
 \frac{4 \pi}{g_b^2} & = & 
\left(1 - \frac{2}{N} \right) \ln \frac{\Lambda^2}{m^2} 
                       + \frac{1}{N} \left(2 \ln \ln \frac{\Lambda^2}{m^2}
\right.\\
&&  \left. -  \mathrm{li}\, \frac{\Lambda^2}{m^2} + 2 \gamma_E + 4 \ln 2
                       \right)\;. 
\label{gapintermsofm}
\eqa
To see that Eq.~(\ref{gapintermsofm}) 
becomes finite after coupling constant renormalization, we
first express it in terms of $m_\phi^2$, using Eq.~(\ref{massshift2}):
\bqa \nonumber
 \frac{4 \pi}{g_b^2} & = & \left(1 - \frac{2}{N} \right) \ln
\frac{\Lambda^2}{m_\phi^2}  
     + \frac{2}{N} \bigg(\ln \ln \frac{\Lambda^2}{m_\phi^2} 
\\ &&
  + \gamma_E + \ln 4 \bigg)\;.
\label{gapintermsofmphi}
\eqa
The renormalization constant for $g$ is denoted by $Z_{g^2}^{-1}$ and
is given by
\bqa\nonumber
Z_{g^2}^{-1} & = & 1+{g^2\over4\pi}
\left(1-{2\over N}\right)\ln{\Lambda^2\over\mu^2}+
{1\over N}{g^2 \over 2\pi}\ln \ln
{\Lambda^2 \over \mu^2} \;,
\\ &&
\label{Z1}
\eqa
Making the substitution $g_b^2 \rightarrow Z_{g^2} g^2(\mu)$, 
we obtain the renormalized gap equation:
\bqa
{4\pi\over g^2(\mu)}&=&
\left(1-{2\over N}\right)\ln{\mu^2\over m_{\phi}^2}+{2\over N}
\left(\gamma_E+\ln4 \right)\;.
\label{gapren}
\eqa
The expression Eq.~(\ref{Z1}) for $Z_{g^2}^{-1}$ is exact in 
$g^2(\mu)$ up to order $1/N^2$ corrections and results in the known NLO 
$\beta$-function~\cite{Rim-Weisberger,Orloff-Brout,BCR}:
\bqa
  \beta(g_b^2) & = &  
  \Lambda {\mathrm{d} g_b^2 \over \mathrm{d}\Lambda} 
= -\left(1-{2\over N}\right)\frac{g_b^4}{2\pi} \left(1 +
\frac{1}{N}\frac{g_b^2}{2\pi} \right)\;, \\
  \beta(g^2)  & = & 
  \mu {\mathrm{d} g^2 \over \mathrm{d}\mu} 
= - \frac{g^4}{2\pi} \left(1 - \frac{2}{N} \right) \;.
\eqa

Using the gap equation, one can obtain 
the value of the effective potential ${\cal V}$ at 
the minimum, where it equals the pressure ${\cal P}$.
In terms of bare quantities, we obtain
\beq
{\cal P}^{T=0}=  -\left(N - 2 \right)\frac{m^2}{8\pi}
     - \frac{1}{8 \pi} \Lambda^2 \ln \frac{4\pi}{g_b^2} \;.
\label{pressure0}
\eeq
This equation will be used to subtract the pressure at zero temperature from
the pressure at finite temperature.

\section{Finite temperature} 
The results at zero temperature are obtained analytically, but at finite 
temperature this is in general not possible.
Therefore, we will investigate the pressure numerically. However, 
we are able to isolate the ultraviolet divergences analytically.

The effective potential through next-to-leading order in $1/N$ is now 
given by 
\bqa\nonumber
{\cal V} &=&
{m^2N\over2g^2}-{1\over2}N\sumint_P\ln\left[P^2+m^2\right]
\\ &&
-{1\over2}\sumint_{P}\ln\left[\Pi^T(P,m)\right]\;,
\label{P-NLO}
\eqa
where the inverse propagator $\Pi^T(P,m)$ is the finite temperature version
of Eq.~(\ref{Pi}) and we have defined the sum-integral
\bqa
  \hbox{$\sum$}\!\!\!\!\!\!\int_{P}& \;\equiv\; &
   T\sum_{p_0=2n\pi T}\:\int {\mathrm{d} p \over 2 \pi}
\;.
\label{sumint-def}
\eqa
Summing over Matsubara frequencies and averaging over angles,
$\Pi^T(P,m)$ reduces to
\bqa
\Pi^T(P,m)&=&
\int_{-\infty}^{\infty}{\mathrm{d} q \over E_q} R(P,q)
\left[1+2n(E_q)\right]\;,
\eqa
where $E_q=\sqrt{q^2+m^2}$ and $n(x)=(\exp(\beta x)-1)^{-1}$ is the
Bose-Einstein distribution. The function $R(P,q)$ is given by
\bqa
R(P,q)&=& - {1\over4\pi}{P^2+2pq\over(P^2+2pq)^2+4p_0^2E_q^2}
\;.
\eqa
The inverse propagator cannot be evaluated analytically, so we will
evaluate it numerically. For this purpose, it is necessary to isolate
ultraviolet 
divergences analytically. As expected on general grounds, i.e.\ from the 
absence of temperature-dependent ultraviolet divergences,  
and as verified numerically, the quantity
\bqa
F_1 & = & \sumint_P \ln \Pi^T(P, m) - \int \spaceint{2}{P} \ln \Pi^T(P, m),
\eqa 
is finite. 
To calculate $F_1$ we used an Abel-Plana formula~\cite{Barton}.
In order to isolate the
divergences we consider the limit $p \gg T$, where we can approximate
\bqa \nonumber
\Pi^T(P,m)&\approx& \Pi^{}(P,m) - \frac{1}{4 \pi} \frac{P^2}{P^4 + 4 m^2
p_0^2} J_1\\
& \equiv & \Pi^T_{\rm HE}(P,m)\;,
\label{PiHE}
\eqa
where $\Pi^{}(P,m)$ is given in Eq.~(\ref{Pi}) and 
\beq
J_1({\beta m}) = 4 \int_0^\infty {\mathrm{d}q\,\over E_q}\, n(E_q)\;.
\eeq
In order to split off the prefactor of the logarithm in $\Pi^{T=0}$, 
we define 
$\tilde \Pi(P, m) = - 4\pi\sqrt{P^2(P^2 + 4m^2)} \Pi(P,m)$.
This gives the following
contribution to the free energy 
\bqa \nonumber 
 D_1 &=& -\frac{1}{2} \int \spaceint{2}{P} \ln \left[P^2 + 4m^2\right] \\
 & = & 
 -\frac{m^2}{2 \pi} \left(1 + \ln \frac{\Lambda^2}{4 m^2} \right).
\eqa
In order to isolate the infinities, we need the large-$P$
behavior of 
$\tilde\Pi^T_{\rm HE}(P,m)$:
\bqa \nonumber
\tilde \Pi^T_{\rm HE}(P,m) &=& \ln {P^2\over \bar{m}^2} + {2m^2\over P^2}
\left(1 + J_1 \right)\\
& & - {4 m^2 p_0^2\over P^4} J_1 + {\cal O}\left({m^4\over P^4}\right),
\eqa
where $\bar{m}^2 = m^2 \exp(-J_1)$. This yields
\bqa
\int {\mathrm{d}^2 P \over (2\pi)^2} 
\ln\left[\tilde \Pi_{\rm HE}^T(P,m)\right] & = & D_2 + {\rm finite} \
{\rm terms}\;,
\eqa
where
\bqa \nonumber
D_2 & = & 
\frac{1}{4 \pi} \left[\Lambda^2 \ln \ln \frac{\Lambda^2}{\bar m^2}
 - \bar m^2 \, \li \frac{\Lambda^2}{\bar m^2} \right] \\
& & + \frac{m^2}{2 \pi} \ln \ln \frac{\Lambda^2}{\bar m^2} \;.
\label{D2}
\eqa
Finally, we define 
\bqa
F_2 & = & \int \spaceint{2}{P} \ln\left[\tilde \Pi(P, m)\right] - D_2 \;.
\eqa
Again we have checked numerically that the quantity $F_2$ is finite,
demonstrating that we have identified all ultraviolet divergences. 
  
Putting everything together, the finite temperature
effective potential becomes
\bqa \nonumber
{\cal V}_{\rm} & = &  \frac{N m^2}{2 g_b^2}
-\frac{N m^2}{8\pi}\left(1 + \ln \frac{\Lambda^2}{m^2}  \right)
  + \frac{N}{8\pi} T^2 J_0
\\
&& - \frac{1}{2} \left(F_1 + D_1 + F_2 + D_2 \right) \;,
\eqa
where 
\beq
J_0(\beta m) = {8\over T^2} \int_0^\infty {\mathrm{d}q\, q^2 \over
E_q} n(E_q)\;.
\eeq
We again note that we have systematically dropped $m$-independent 
divergences and terms that vanish in the limit $\Lambda^2 \to \infty$.

The gap equation~(\ref{gap}) at nonzero temperature now becomes 
\bqa\nonumber
  \frac{4 \pi}{g_b^2} & = & \ln \frac{\Lambda^2}{\bar m^2}                  
  + \frac{1}{N} \left[2 \ln \ln \frac{\Lambda^2}{\bar m^2}
          -  \frac{\mathrm{d} \bar m^2}{\mathrm{d} m^2}
      \li \frac{\Lambda^2}{\bar m^2} \right.\\
&& \left. - 2 \ln \frac{\Lambda^2}{4 m^2} + 
          4 \pi \frac{\mathrm{d}(F_1+F_2)}{\mathrm{d} m^2} \right] \;.
\label{g2}
\eqa
From the fact that $g_b^2$ is temperature independent, one can conclude that
$\bar{m}^2$ is also temperature independent at leading order in the $1/N$
expansion, when it is a solution to the gap equation. We will use this fact 
later on to conclude that the pressure can be renormalized in a 
temperature-independent way.  

The calculation of the self-energy $\Sigma(P,m)$ at $p_0=0$ and
$p^2=-m_\phi^2$, at finite temperature yields
\beq
  m^2 = m_\phi^2 - \frac{\bar m^2_\phi}{N} \left[ \li 
\frac{\Lambda^2}{\bar m_\phi^2} + F_3\right]\;,
  \label{massshiftnonzerotemp}
\eeq
where $F_3$ is a finite function that depends on 
the temperature as well as $m_\phi$.
Since we use $m_\phi$ merely as a way to express the renormalized gap 
equation in terms of finite quantities, any choice of $F_3$ will do and we
choose $F_3=0$. We have checked numerically that other choices indeed do not
alter the final result for the pressure. 
Using Eq.~(\ref{massshiftnonzerotemp}), Eq.~(\ref{g2}) now becomes
\bqa\nonumber
  \frac{4 \pi}{g_b^2} & = & \ln \frac{\Lambda^2}{\bar m_\phi^2}                
  + \frac{1}{N} \left[2 \ln \ln \frac{\Lambda^2}{\bar m_\phi^2}
        \right.\\
 & & \left.  - 2 \ln \frac{\Lambda^2}{4 m_\phi^2} +
          4 \pi \frac{\mathrm{d}(F_1+F_2)}{\mathrm{d} m_\phi^2} \right] \;.
\eqa
To render the gap equation finite, we again only need to make the 
substitution $g_b^2 \rightarrow Z_{g^2} g^2(\mu)$, where $Z_{g^2}^{-1}$ is
given by Eq.~(\ref{Z1}). The renormalized gap equation then becomes
\bqa\nonumber
  \frac{4 \pi}{g^2(\mu)} & = & \left(1-{2\over N} \right) 
\ln \frac{\mu^2}{\bar m_\phi^2}\\
&& \mbox{} \hspace{-1 cm} 
+ \frac{2}{N} \left[J_1(\beta m_\phi)+ \ln 4 + 2 \pi 
{\mathrm{d}(F_1+F_2)\over \mathrm{d} m_\phi^2} \right] \;.
\label{RenGapEqNonzeroT}
\eqa

\begin{figure}[htb]
\epsfysize=5.2cm
\epsffile{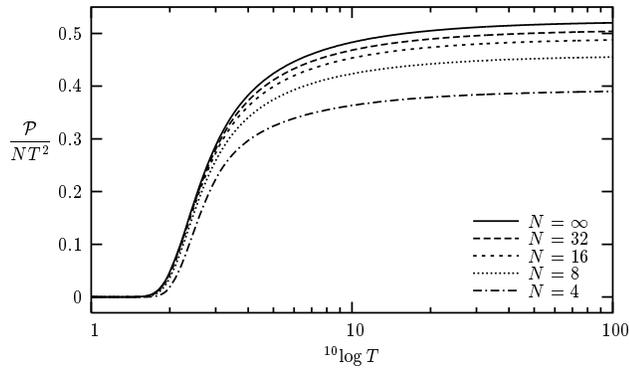}
\caption[a]{Pressure as a function of temperature at NLO for different values
of $N$.} 
\label{s12}
\end{figure}

Using the gap equation, we obtain the value of the effective potential at the
minimum which is equal to the pressure. Using 
Eq.~(\ref{massshiftnonzerotemp}) and
expanding the $J_0$ and $J_1$ functions, one ultimately obtains for 
the pressure at nonzero temperature minus the pressure Eq.~(\ref{pressure0}) 
at zero temperature, 
\bqa\nonumber
{\cal P} \equiv {\cal P}^{T} - {\cal P}^{T=0} 
& = & \frac{N-2}{8 \pi} \left[ m_{\phi}^2(0)
        -m_\phi^2 \right] \\
\nonumber
& + & \frac{N}{8\pi} \left[ T^2 J_0(\beta m_\phi) +  m_\phi^2 
J_1(\beta m_\phi) \right] \\
&+ & \frac{1}{2} \left[m^2_\phi {\mathrm{d}(F_1+F_2)\over \mathrm{d}
m_\phi^2} - F_1 - F_2 \right] \;,
\eqa
where $F_1$ and $F_2$ are functions of $T^2$ and $m_\phi^2 = m_\phi^2(T)$, 
and $m_{\phi}^2(0) = m_\phi^2(T=0)$. 
We have numerically evaluated the expression for the 
pressure, after solving Eq.~(\ref{RenGapEqNonzeroT}) for $m_\phi(T)$.
The result for different values of $N$ is shown in Fig.\ \ref{s12}, for the
arbitrary choice $g^2(\mu=500) = 10$, hence $T$ is given in the same units as
$\mu$. 
As can be shown and seen in the figure, ${\cal P}/N T^2$ approaches an 
$N$-dependent constant (to be evaluated below) at large temperatures, which
for $N \to \infty$ is $\pi/6$. Moreover, it approaches zero 
in the limit of zero temperature. If we normalize the pressure,
for a given value of $N$, to its value at $T = \infty$, we find that
the normalized pressure has a very small dependence on $N$.  

\section{High-temperature approximations}
In Ref.~\cite{kap}, Bochkarev and Kapusta consider 
the nonlinear sigma model in 3+1 dimensions, 
which is nonrenormalizable, at NLO 
in the $1/N$ expansion. Since the result
for the pressure cannot be obtained analytically, they resort to a
``high-energy approximation''. We will make the same approximation
and compare it with the exact numerical results obtained in Sec.~III.

The idea of the high-energy approximation is that
in the part of $\Pi^T$ proportional to the distribution function $n(E_q)$, 
the important
contribution comes from the region where $p_0,p \gg q$.
One can therefore approximate the self-energy $\Pi^T(P,m)$
by its high-energy behavior.  
In the present case, this amounts to 
\bqa\nonumber
\Pi^T(P,m)&\approx& \Pi(P,m) - \frac{1}{4 \pi} \frac{P^2}{(p_0^2
+\omega_+^2) (p_0^2 + \omega_-^2)} J_1\;,
\\ &&
\eqa
where $\omega_{\pm}=\sqrt{p^2+m^2}\pm m$. 
This expression is identical to 
Eq.~(\ref{PiHE}). After simply discarding the $T=0$
contribution to $\Pi^T(P,m)$, as done in Ref.~\cite{kap}, the effective
potential is approximately given by 
\bqa\nonumber
{\cal V}_{\rm HEA}& =&
{m^2N\over2g^{2}}-{1\over2}N\sumint_{P}\ln\left[P^2+m^2\right]
-{1\over2}\sumint_{P}\ln P^2
\\&&
\hspace{-1cm}
+{1\over2}\sumint_{P}\ln\left[p_0^2+\omega_{+}^2\right]
+{1\over2}\sumint_{P}\ln\left[p_0^2+\omega_{-}^2\right]
\;.
\label{pnr}
\eqa
The resulting expression for the gap equation is
\bqa\nonumber
N g^{-2}&=&N\sumint_{P}{1\over P^2+m^2}
-\sumint_{P}{\omega_{+}^2\over mE_p}{1\over p_0^2+\omega_{+}^2}
\\
&&
+\sumint_{P}{\omega_{-}^2\over mE_p}{1\over p_0^2+\omega_{-}^2}
\;.
\label{gnlo}
\eqa
Again, the gap equation requires coupling
constant renormalization. In this approximation, the
renormalization constant is
\bqa 
Z_{g^2}^{-1}=1+{g^2\over4\pi} \left(1-{2\over N}\right) 
\ln{\Lambda^2\over\mu^2}\;,
\eqa 
which is 
consistent with the perturbative renormalization constant to leading order 
in $g^2$. 
Making the substitution $g_b^2\rightarrow Z_{g^2}g^2$, we obtain
\bqa
\mbox{} \hspace{-0.5 cm} 1&=&{g^2\over4\pi N}\left[NJ_1-K_1^{+}-K_1^{-}
+2(N-2)\ln{\mu\over m}
\right]\;.
\eqa
where the function $K_1^{\pm}$ is
\bqa
K_1^{\pm} & = & \pm 4\int_0^{\infty}
{\mathrm{d}p\;\omega_{\pm}\over mE_p}
n(\omega_{\pm})\;. 
\eqa
Note, however, that
the pressure is finite even when we substitute the
unrenormalized gap equation~(\ref{gnlo}) into Eq.~(\ref{pnr}): 
\bqa\nonumber
{\cal P}_{\rm HEA}&=&
{N\over8\pi}\left[J_0T^2+(J_1-1)m^2
\right]+{\pi\over6}T^2
\\
&& \mbox{} \hspace{-2 cm}
-{1\over8\pi}\left[(K_0^{+}+K_0^{-})T^2+(K_1^{+}+K_1^--2)m^2\right]
\;,
\label{pnlo}
\eqa
where the function $K_0^{\pm}$ is
\bqa
K_0^{\pm} & = & {8\over T^2}
\int_0^{\infty}{\mathrm{d}p\;p^{2}\over E_p}n(\omega_{\pm})\;.
\eqa

\begin{figure}[htb]
\epsfysize=5.2cm
\epsffile{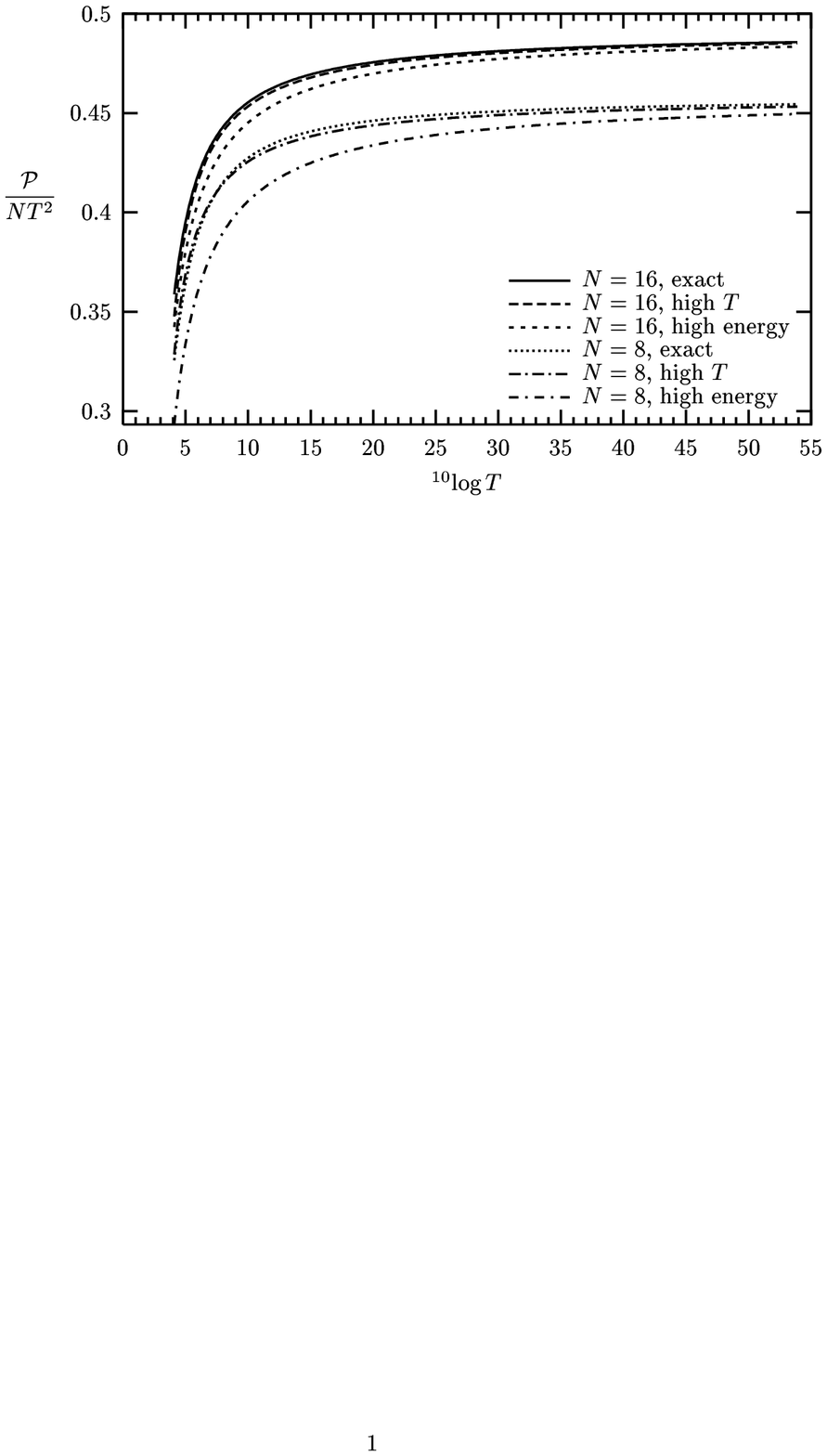}
\caption[a]{Pressure as a function of temperature at NLO for different
values of $N$ compared with the high-energy and high-temperature 
approximations.}
\label{s13}
\end{figure}
From Fig.~\ref{s13}, one can see that the high-energy 
approximation underestimates the pressure compared to the exact result. 
The advantage of an approximation like the high-energy approximation is that
the analytic calculations are simpler and that 
it is easier to implement numerically. 

We suggest a different approximation, which is better than the high-energy
approximation. We will calculate the inverse $\tilde \alpha$-propagator 
$\Pi^T$ by first integrating over the momentum. We obtain
\bqa \nonumber
  \Pi^T(P, m) & = & -  \frac{1}{2 \beta} 
\sum_{q_0=2n\pi T}
  \frac{1}{\sqrt{m^2 + q_0^2}} \\
& & \times \frac{P^2 + 2 q_0 p_0}
  {P^4 + 4 q_0 (q_0 + p_0) P^2 + 4m^2 p^2} \;.
\label{pipi}
\eqa
In the limit $m \ll T$, we can approximate $\Pi^T(P, m)$ by 
$\Pi_{\mathrm{HT}}^T (P,m)$, where
keep only the $q_0 = 0$ mode in the sum Eq.~(\ref{pipi}):
\bqa
\Pi_{\mathrm{HT}}^T (P,m)&=& 
-\frac{1}{2} \frac{1}{\beta m} \frac{P^2}{P^4 + 4 m^2 p^2}\;.
\eqa
Since it follows from the leading order 
gap equation that for high temperature and for all values of the 
coupling constant, 
$m \ll T$, we call this approximation the high-temperature (HT) 
approximation. 
By using that $P^4 + 4m^2 p^2$ can be written as 
$[p_0^2 + (p+ im)^2 + m^2][p_0^2 + (p-im)^2 +m^2]$
and shifting $p \rightarrow p \pm im$ after taking the logarithm, 
the functions $F_1$ and $F_2$ can be approximated by
\begin{equation}
  F_1 \approx \frac{1}{2\pi} T^2 J_0(\beta m) - \frac{\pi}{3}T^2\;,  \;\;\;\;
  F_2 \approx 0 \;.
\label{ApproxofFs}
\end{equation}
A numerical calculation of $F_1 + F_2$ shows that for $m/T \simordertwo 0.1$ 
this approximation has an error smaller than $10$ percent. 
Approximating $F_1$ and $F_2$ using the high-energy
approximation is less accurate. The result for the pressure in the 
high-temperature approximation is shown for comparison
in Fig.~\ref{s13} (again
for $g^2(\mu=500) = 10$). 

One can approximate the pressure even further by expanding the functions $J_0$
and $J_1$ in the limit $\beta m\rightarrow 0$:
\bqa\nonumber
J_0&=&{4\pi^2\over3}-4\pi\beta m
-2\left(\log{\beta m\over4\pi}+\gamma_E-{1\over2}\right)(\beta m)^2 \\
&&+{\cal O}\left((\beta m)^4\right) \;, \\
J_1&=&{2\pi\over\beta m}
+2\left(\log{\beta m\over4\pi}+\gamma_E\right) + {\cal O}\left((\beta
m)^2\right)\;.
\label{ApproxofJ1}
\eqa
Inserting the approximations given in Eqs.\ (\ref{ApproxofFs}) and 
(\ref{ApproxofJ1}) into the gap equation~(\ref{RenGapEqNonzeroT}), 
one obtains
\beq
\beta m \approx \pi \left[ \left({2\pi\over g^2(\mu)}-\frac{\ln
4}{N}\right)\left(1+{2\over N} \right)-\gamma_E - \ln {\mu\beta \over 4\pi}
\right]^{-1}\;,
\eeq
which indicates that $\beta m \sim 1/\ln T$ for large $T$. 
In the limit $m/T \rightarrow 0$, we 
obtain for the high-temperature approximation of the pressure
\begin{equation}
  \frac{\mathcal{P}}{N T^2} \approx \frac{\pi}{6} \left(1 - \frac{1}{N}\right)
  - \left(1-{2\over N}\right) {m \over 4 T}\;, 
\end{equation}
where the first term is the pressure of a gas of free massless 
particles with $N-1$ degrees of freedom.

\section{Thermal renormalons}

We have shown that a finite pressure at finite temperature 
can be obtained after subtraction of the zero-temperature pressure and 
coupling constant renormalization.
This agrees with the general expectation that 
ultraviolet divergences
are connected with short-distance physics and 
therefore independent of the temperature. While we have shown this
explicitly at NLO in the $1/N$ expansion, this is not 
the case for the effective potential away from its minimum. 
 
\begin{figure}[htb]
\epsfysize=2.6cm
\epsffile{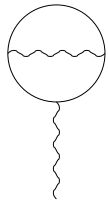}
\caption[a]{Tadpole diagram contributing to $\langle\alpha\rangle$
at next-to-leading order in $1/N$. The wavy line represents the $\tilde
\alpha$ propagator and the solid line the $\Phi$ propagator.}
\label{tadpole}
\end{figure}

In the expression Eq.~(\ref{intlogPi}) for the effective potential 
at zero temperature, the two contributions
$\Lambda^2 \ln \ln (\Lambda^2/m^2)$ and $m^2 {\rm li}(\Lambda^2/m^2)$ 
cannot be removed using $m$-independent 
counterterms~\footnote{Note that the quantity 
$\Lambda^2 \ln \ln \frac{\Lambda^2}{m^2}- 
\Lambda^2 \ln \ln \frac{\Lambda^2}{\mu^2}$ (with $\mu \neq m$) diverges as 
$\Lambda^2 \to \infty$, whereas $\ln \ln \frac{\Lambda^2}{m^2}-\ln \ln 
\frac{\Lambda^2}{\mu^2}$ vanishes.}. 
While this may not be a problem at zero temperature, it would
certainly become one at finite temperature when $m$ becomes 
a function of
temperature. This would imply temperature-dependent renormalization, which
is not acceptable. In Ref.~\cite{BCR}, these two divergences are dealt with by
considering the effective potential normalized to its zero-mass value, i.e.
${\cal V}_{\rm NLO}(m) - {\cal V}_{\rm NLO}(0)$. This subtraction is
ill-defined due to infrared divergences and 
therefore one should understand it as subtracting the contributions from 
$\ln\left[\Pi(P,m)\right]$
obtained in the limit $P^2/m^2 \to \infty$~\cite{CR}. This is called the 
``perturbative tail''.
If we denote $\Pi(P,m)$ in this limit by $\Pi_\infty(P,m)$, one finds
$\Pi_{\infty}(P,m)= \ln(P^2/m^2)/(4\pi P^2)$ and 
\bqa\nonumber
\int
\frac{\mathrm{d}^2P}{(2\pi)^2}\ln\left[\Pi_\infty(P,m) \right]& = &  
\frac{1}{4\pi} 
\left[ \Lambda^2 \ln \ln \frac{\Lambda^2}{m^2} 
\right.
\\ && \mbox{} \hspace{-1cm}
\left.
- m^2 {\rm li}\left(\frac{\Lambda^2}{m^2} \right)
\right]\;,
\eqa
where we have implicitly used the principal-value prescription. 
In Refs.~\cite{BCR,CR,CR-90}, this subtraction is not motivated, but we 
point out
that the subtracted contribution is associated with IR renormalons. 
As shown in~\cite{David-86},
the vacuum expectation value of $\alpha$, i.e.\ $m^2$, is 
inherently ambiguous, when one tries to separate (in order to subtract) 
perturbative contributions proportional 
to $\Lambda^2$ from the nonperturbative ones proportional to $m^2$ in the 
limit where
$\Lambda^2 \gg m^2$. In~\cite{David-86}, it was shown that
\beq
\amp{0|\alpha|0}  =  m_{\rm LO}^2 + \frac{4\pi m_{\rm LO}^2}{N} 
\int \frac{\mathrm{d}^2P}{(2\pi)^2} \frac{1}{\Pi} \frac{\partial \Pi}{\partial 
m_{\rm LO}^2} + {\cal O} \left(\frac{1}{N^2}
\right)\;,
\label{vevalpha}
\eeq
where $m_{\rm LO}^2\equiv \Lambda^2 \exp (-4\pi/g_b^2)$ and the $1/N$ 
contribution arises from the tadpole diagram shown in Fig.~\ref{tadpole}

One can show that this equation is in agreement with 
the gap equation~(\ref{gapintermsofm}) 
if we write $m^2 = m_{\rm LO}^2 +
m_{\rm NLO}^2/N$.  

The part of the integral in Eq.\ (\ref{vevalpha}) 
that has the IR renormalon pole in the Borel plane 
is in fact the contribution from the integrand in the limit 
$P^2/m^2 \to \infty$:
\bqa \nonumber 
\int^\Lambda \frac{\mathrm{d}^2 P}{(2\pi)^2} \frac{1}{\Pi_\infty} \frac{\partial
\Pi_\infty}{\partial m^2} & = & - \frac{1}{4\pi} 
{\rm li}\left({\Lambda^2 \over m^2}\right)
\\
&= & 
- \frac{1}{4\pi} \frac{\Lambda^2}{m^2} e^{-x} {\rm Ei} (x)\;,
\label{renormalon}
\eqa
where $x=\ln(\Lambda^2/m^2)$. In the limit 
$x\to \infty$, the logarithmic integral has the asymptotic expansion: 
\bqa \nonumber 
e^{-x} {\rm Ei} (x) & = & \sum_{n=0}^\infty \frac{n!}{x^{n+1}} \mp i\pi
e^{-x} \\
& = & \int_0^\infty \mathrm{d}b \, \frac{e^{-bx}}{1-b} \mp i\pi
e^{-x} \;,
\label{rp}
\eqa
where $\arg(b) = \pm \varepsilon$. 
From Eq.~(\ref{rp}), it is clear that there is a renormalon
pole at $b=1$. 
This shows that when $\Lambda \to
\infty$ the value of 
$\amp{0|\alpha|0}$ is inherently ambiguous at NLO, due to the freedom in the
choice of prescription. David has shown that this ambiguity also arises in
dimensional regularization~\cite{David-84}.  

The same problem appears in the calculation of the
effective potential, but {\em not\/} in the gap equation. The latter can be 
seen from the last term of Eq.\ (\ref{V0}) which contributes to the gap 
equation as follows
\bqa 
{1\over 2} \frac{\partial}{\partial m^2} \int \frac{\mathrm{d}^2
P}{(2\pi)^2}\ln \Pi  = {1\over 2} \int \frac{\mathrm{d}^2
P}{(2\pi)^2} \frac{1}{\Pi} \frac{\partial \Pi}{\partial m^2} \;.
\eqa 
The ambiguity that would arise from this term when removing its perturbative
tail (cf.\ Eq.\ (\ref{renormalon})) cancels in the gap equation
(\ref{gapintermsofm}) against the one arising in 
$m^2$ (cf.\ Eq.\ (\ref{vevalpha})).

The perturbative tail of the 
effective potential, i.e.\ the first two terms of $D_2$ defined
in Eq.~(\ref{D2}),
corresponds to poles in the Borel plane at $b=0$ and $b=1$, respectively.
Since $\bar m^2$
is only temperature independent at the minimum (at LO only, but that is
sufficient since we are working at NLO), the subtraction of the
perturbative tail will become temperature dependent, {\em except\/} at the 
minimum. Since subtracting temperature-dependent divergences renders the
remaining temperature-dependent terms ambiguous, 
we refrain from following this
strategy and thus from trying to define a finite effective potential at finite
temperature. In order to avoid any renormalon ambiguity, we have also
not considered obtaining a finite effective potential or even a finite 
pressure at zero temperature.
However, we have calculated the quantity 
${\cal P}^{T} - {\cal P}^{T=0}$, which is free of renormalon ambiguities 
and is finite after 
temperature-independent coupling constant renormalization

Finally, we comment on 
the possible temperature dependence of renormalon
contributions to $\langle\alpha\rangle$ and the effective potential.
One can show that 
Eq.~(\ref{vevalpha}) at finite temperature has exactly the same renormalon
contribution, i.e.\ neither the pole nor the residue become temperature
dependent. Secondly, the perturbative tail of the 
effective potential which is given by the first two terms of $D_2$,
corresponds to poles in the Borel plane at $b=0$ and $b=1$. 
The positions of the renormalon poles are not affected by temperature.
Only the residues become temperature dependent, 
except at the minimum of the 
potential, as we concluded earlier. 
The fact that renormalon pole positions are not affected 
by temperature, but residues are, is also the case
for the thermal ultraviolet renormalons 
in $\phi^4$ in $3+1$ dimensions
studied by the authors of Ref.~\cite{Loewe}.

\section{Summary and Conclusions}
 
To summarize, we have calculated the pressure in the NLSM at 
finite temperature to NLO in the $1/N$ expansion. Our main result is that we 
obtain an unambiguous, finite pressure, 
by subtracting the zero-temperature value of the pressure
and renormalization of the coupling 
constant in a temperature-independent way.
This procedure cannot be carried out away from the
minimum of the effective potential and we have 
argued that defining a finite, effective potential by the subtraction of the
so-called perturbative tail, leads to ambiguities associated with
IR renormalons.
In general, these become temperature dependent, and this casts
doubt on the usefulness of defining a finite effective potential. 

We have calculated the expression for the pressure at finite temperature 
numerically and observe
that the $1/N$ expansion is a meaningful expansion for all temperatures.
We have also investigated the high-energy approximation
that was originally  
applied to the NLSM in 3+1 dimensions by Bochkarev and Kapusta.
In 1+1 dimension, where one can compare with exact numerical results, 
we have shown that it underestimates the pressure for all temperatures. We
have suggested an improved approximation, the so-called high-temperature
approximation. This approximation has the advantage that it is quite easy to 
produce numerical results and agrees better with the exact results. 
At asymptotically high temperatures the pressure approaches that of a gas of 
$N-1$ free massless particles. 

\section*{Acknowledgments} 

We would like to thank Jan Smit for useful discussions. H.J.W.~thanks Rob
Pisarski for a fruitful discussion. 
The research of D.B.~has been made possible by financial support from the 
Royal Netherlands Academy of Arts and Sciences.

\end{document}